\documentclass[aps, pre, twocolumn,showkeys,showpacs,amsmath,amssymb,superscriptaddress]{revtex4-1}

\usepackage[T1]{fontenc}
\usepackage[latin1]{inputenc}

\usepackage[ngerman,british]{babel}
	\addto\captionsngerman{}

\usepackage{amsmath,amssymb,amsthm,dsfont,mathtools,mathrsfs,bbm}

\usepackage{mdwlist}
\usepackage{paralist}
\usepackage{array}
\usepackage{booktabs}						
\usepackage{dcolumn} 						

\usepackage{graphicx}
\usepackage{wrapfig}

\usepackage{url}

\usepackage{color}
	\definecolor{myblue}{rgb}{0,0.3,0.8}
	\definecolor{mygreen}{rgb}{0,0.5,0}
	\definecolor{myblue}{rgb}{0,0.3,0.8}

\usepackage[version=3]{mhchem}

\usepackage{savesym}

\savesymbol{a}
\savesymbol{b}
\savesymbol{d}
\savesymbol{e}
\savesymbol{k}
\savesymbol{l}
\savesymbol{H}
\savesymbol{L}
\savesymbol{p}
\savesymbol{r}
\savesymbol{s}
\savesymbol{S}
\savesymbol{F}
\savesymbol{t}
\savesymbol{w}
\savesymbol{th}
\savesymbol{div}
\savesymbol{tr}
\savesymbol{Re}
\savesymbol{Im}
\savesymbol{AA}
\savesymbol{SS}
\savesymbol{Si}
\savesymbol{gg}
\savesymbol{ss}

\newcommand{\a}{\alpha}
\newcommand{\b}{\beta}
\newcommand{\e}{\varepsilon}
\newcommand{\d}{\delta}
\newcommand{\del}{\partial}

\newcommand{\g}{\gamma}
\newcommand{\G}{\Gamma}

\newcommand{\r}{\rho}
\newcommand{\s}{\sigma}

\newcommand{\w}{\omega}

\newcommand{\dt}{{\Delta t}}

\newcommand{\cov}{{\nabla}}

\newcommand{\R}{\mathbbm{R}}
\newcommand{\C}{\mathbbm{C}}

\DeclareMathOperator{\tr}{Tr}

\newcommand{\T}{\textstyle}

\newlength{\tmplen}

\newcommand{\norm}[1]{\left\lVert#1\right\rVert^2}

\usepackage{bbold}

\begin{document}

\title[AC]{Quantized alternate current on curved graphene}

\author{Kyriakos Flouris}
\affiliation{ %
ETH
 Z\"urich, Computational Physics for Engineering Materials, Institute
 for Building Materials, Wolfgang-Pauli-Str. 27, HIT, CH-8093 Z\"urich
 (Switzerland)}%
\author{Sauro Succi}
\affiliation{ %
Center for Life Nano Sciences at La Sapienza, Istituto Italiano di Tecnologia, viale R. Margherita, 265, 00161, Roma, Italy}%
 \affiliation{ %
Institute for Applied Computational Science, J. Paulson School of Engineering and Applied Sciences, Harvard University, 29 Oxford Street, Cambridge, USA}%
\author{Hans J. Herrmann}
\affiliation{ %
Departamento de F\' isica, Universidade do Cear\' a, 60451-970 Fortaleza, Brazil}%
\affiliation{ %
C.N.R.S., UMR 7636, PMMH, ESPCI, 10 rue Vauquelin, 75231 Paris Cedex 05, France }%

\begin{abstract}%
Based on the numerical solution of the quantum lattice Boltzmann method in curved space, we predict the onset 
of a quantized alternating current on curved graphene sheets.  
Such numerical prediction is verified analytically via a set of semi-classical equations relating 
the Berry curvature to real space curvature.  
The proposed quantised oscillating current on curved graphene could form the basis for 
the implementation of quantum information processing algorithms. 
\end{abstract}

\maketitle

\section{Introduction}

In  recent years, the most puzzling features of quantum mechanics, such as entanglement and 
non-local "spooky action at distance", long regarded as a sort of extravagant speculations, have 
received spectacular experimental confirmation  \cite{EPR,bell,Aspect,Haroche}. 
Besides their deep fundamental implications, such phenomena may also open up
transformative scenarios for material science and related applications 
in quantum computing and telecommunications \cite{bennett,Zeilinger}.
Along with such burst of experimental activity, a corresponding upsurge of theoretical and computational
methods has also emerged in the last two decades, including, among others, new quantum-many body 
techniques, quantum simulators \cite{cirac,blochImanuel,quantumsim,quantization} and
quantum walks \cite{quantumenta}.

Quantum walks were first introduced by Aharonov and collaborators in 1993 \cite{Aharanov}, just a few months before the appearance of the first quantum lattice Boltzmann scheme
\cite{succi_qlbm}, which was only recently recognized to be a quantum walk too \cite{succiQW}.
Quantum walks \cite{quantumwalk2,quantumwalk3} are currently utilized to investigate exotic 
states of quantum matter \cite{majorana} and to design new materials
and technologies for quantum engineering applications \cite{quantumwalk}. 

Quantum walks can also help exploring the emergence of classical behaviour in the limit of a vanishing De Broglie length \cite{quantummechanics}. 
Likewise, quantum cellular automata \cite{cellularautomata1,cellularautomata2,cellularautomata3}, can 
be used for simulating complex systems in analogy with their classical counterparts.

Finally, quantum walks have also shown connections with topological aspects of quantum 
mechanics, most notably the Berry phase \cite{Berry}.
Indeed, Berry connection and Berry curvature can be understood as a local gauge potential and gauge 
field, respectively and they define a Berry phase as introduced in 1984 \cite{Berry}. 
The Berry phase has important implications as an analytic tool in topological phases of matter \cite{topologicalstatesofmatter} and, under suitable conditions, it can also be related to real space curvature \cite{ryder}, thus providing a potential bridge between the classical and quantum descriptions of a given system.
 
As of quantum materials, graphene presents one of the most promising
cases for realizing a new generation of quantum devices \cite{graphenerev1,graphenerev2, graphenerev3}. Indeed, since its discovery \cite{Geim-graphene}, this flatland wonder-material has not ceased to surprise scientists with its amazing mechanical and electronic behaviour. For example, stacking graphene sheets at specific angles has shown spectacular
indications of superconductivity and other exotic properties \cite{magic_angle}. 

Tunable transport properties are a basic requirement in electronic devices and specifically in graphene \cite{tunable}. 
Furthermore, it has been shown that graphene sheets can be curved in such a way as to trap particles \cite{cdp}, thus opening further prospects for technological applications based on localized quantum states.

In this work, we propose the generation of a quantized oscillating current on curved graphene, which could be used in conjunction with trapped fermions for the realization of quantum cellular automata. 

Electron transport is simulated by numerically solving the Dirac equation in curved space \cite{JD_thesis, cdp} using an extension to curved space of the   quantum lattice Boltzmann  method \cite{succi_qlbm}. In addition, a simpler representation of the system is solved analytically through a set 
of semi-classical equations of motion, relating Berry to real space curvature.
 
The paper is organized as follows. First, we introduce the Dirac equation and its extension to curved space and specifically deformed graphene. In the subsequent section, we present the results of numerical simulations and finally we conclude with a summary and outlook section. A detailed description of the numerical model is provided in the Appendix (Appendix~\ref{sec:QLBM}).


\section{The Dirac equation in curved space and graphene}

The Dirac equation in curved space can be written in compact notation as follows:
\begin{equation}
\label{eq:Dirac}
(i \g^{\mu}D_{\mu}-m)\Psi=0,
\end{equation}
in natural units $\hbar=c=1$, where $m$ is the particle rest mass, the index $\mu={0,1,2}$ runs over 2D space-time.
In the above, $\Psi = (\Psi^+, \Psi^-) = (\psi_1^+,\psi_2^-,\psi_1^-,\psi_2^+) \in \C^4$ denotes 
the Dirac four-spinor, and $\g^\mu = \g^\a e_\a^{\ \mu}$ are the generalized $\g$-matrices, 
where $\g^\a \in \C^{4\times 4}$ are the standard $\g$-matrices (in Dirac representation). 
The symbol $e_\a^{\ \mu} $ is the tetrad (first index: flat Minkowski, second index: curved space-time). 

Here, the tetrad is defined by $e_{\alpha}^{\mu} g_{\mu \nu}e_{\beta}^{\nu}=\eta_{\alpha \beta}$ \cite{kaku1993quantum},
where $g_{\mu \nu}$ denotes the metric tensor and $\eta_{\alpha \beta}$ is the 
Minkowski metric. The tetrad basis is chosen such that the standard Dirac matrices can be utilised 
with no need to transform to a new coordinate basis. 
The symbol $D_{\mu}$ denotes the covariant spinor derivative, defined as 
$D_{\mu} \Psi=\del_{\mu} \Psi + \G_{\mu} \Psi$, where $\G_{\mu}$ denotes the 
spin connection matrices given by $\Gamma_\mu = - \frac{i}{4} \w_\mu^{\a\b} \s_{\a\b},$ where $\s_{\a\b} = \frac{i}{2} [\g_\a,\g_\b]$ and $\w_\mu^{\a\b}=\quad e_\nu^\a \cov_\mu e^{\nu \b}$
The Dirac equation in curved space describes quantum relativistic Dirac particles (e.g. electrons ) 
moving on arbitrary manifold trajectories. 

The covariant derivative ensures the independence of the Dirac equation of the coordinate basis. The covariance is satisfied by the connection coefficients which can be interpreted physically as a vector potential. The Poincare symmetries are obeyed by the Dirac equation ensuring the special relativistic nature of the wavefunctions.   The mass term represents the Minkowski metric invariant rest mass.  Interactions add to an effective mass by the very definition of covariant derivative,
which places the vector potential on the same mathematical basis as a physical mass. Graphene is modeled by a mass-less Dirac Hamiltonian.

\subsection{Theory of strained graphene}

Using the tight binding Hamiltonian to describe the bi-partite lattice of graphene, it is established 
that in the low-energy limit, the dispersion relation is linear, as described by 
the Dirac cones at the corners of the first Brillouin zone, which can be described by 
the following Dirac Hamiltonian:
\begin{equation}
\label{eq:dirac_hamiltonian}
H_D=-i \int \Psi^\dagger \gamma^0 \gamma^i \del_i \Psi d^2x, 
\end{equation}
in natural units, where $\Psi$ is in the chiral representation. 

In the context of graphene, the general Dirac spinor is defined as 
$\Psi=(\Psi_a^K,\Psi_a^{K'})=(\psi_A^K,\psi_B^{K'},\psi_A^{K'},\psi_B^{K})$, for 
sub-lattices $A,B$ and valleys $K,K'$. 

The equation of motion stemming from this Hamiltonian is precisely the  Dirac equation. 

In this work, we consider a static space-time metric, with trivial time components
\begin{align*}
	g_{\mu\nu} = 
	\begin{pmatrix}
		1 & 0 \\
		0 & -g_{ij}	
	\end{pmatrix},
\end{align*}
where the latin indices run over the spatial dimensions. 

This simplifies the Dirac equation Eq.~(\ref{eq:Dirac}) to
\begin{equation}
\del_t \Psi + \s^a e_a^{~i}(\del_i + \G_i)\Psi = 0-i\g^0 m \Psi,
\end{equation}
with $\s^a = \g^0 \g^a$. After addition of external vector and scalar potentials $A_i(x)$ and $V(x)$ respectively, 
as explained in Ref.~\cite{jd_paper}, the Dirac equation takes the following form:
\begin{equation}
\label{eq:Diracfull}
\del_t \Psi + \s^a e_a^{~i}(\del_i + \G_i- i A_i)\Psi =-i\g^0 (m-V) \Psi.
\end{equation}
Defining the Dirac current as $J^\mu = \overline \Psi \g^\mu \Psi$, the charge density conservation law 
can be written as $\del_t \rho + \nabla_i J^i = 0$, where
$\r = \Psi^\dagger \Psi \in \R$ and the $J^i = \overline{\Psi}\gamma^i \Psi \in \R$.

The standard Dirac Hamiltonian for Eq.~(\ref{eq:Diracfull}) equation is given by:
\begin{equation}
\label{eq:hamiltoniandirac}
H_D=-i \int \Psi^\dagger \sigma^a e_a^{~i}( \del_i + \G_i-i A_i)\Psi \sqrt{g} d^2x,
\end{equation}

For the case of graphene, the effective Hamiltonian reads as follows \cite{OLIVALEYVA}:
\begin{equation}
\label{eq:hamiltoniangraphene}
H^*_D=-i \int \Psi^\dagger \sigma^a (v_a^{*i} \del_i + \G_a^*-i A^*_a)\Psi d^2x,
\end{equation}
where $v_a^{* i}=\d_{a i} + u_{a i} -\beta \e_{a i}$ is the space-dependent 
Fermi velocity, $\G^*_a=\frac{1}{2}\del_j v_a^{* j}$ is a complex gauge vector field 
which guarantees the hermicity of the Hamiltonian and $A^*_a$ is a strain-induced pseudo-vector 
potential, given by $A^*_a=(A_x^*,A_y^*)=\frac{\beta}{2a}(\e_{xx}-\e_{yy},-2\e_{xy}$). 
Furthermore, $\beta$ is the material-dependent electron Grueneisen parameter, $a$ the lattice spacing 
and $\e_{i\jmath}= u_{i\jmath} +\frac{1}{2}\del_i h \del_j h$ is the general strain tensor, with 
in-plane, $u_{i\jmath}$ and out of plane, $h$ deformations. 

Comparing this to the standard Dirac Hamiltonian in curved space Eq.~(\ref{eq:hamiltoniandirac}), we 
can match both Hamiltonians $H_D$ and $H^*_D$ by fulfilling the following relations:
\begin{equation}
\label{eq:effectivefields}
v_a^{*i}= \sqrt{g}e_a^{~i}, \ \ \G_a^*=  \sqrt{g}e_a^{~i}\G_i, \ \ A^*_a=  \sqrt{g}e_a^{~i}A_i.
\end{equation}

All three relations above can be simultaneously fulfilled by an effective metric tensor 
derived from the explicit expression of the tetrad \cite{jd_paper}.  

The numerical solutions are obtained with the Quantum Lattice Boltzmann Method, as 
described in Appendix~\ref{sec:QLBM} and Ref.~\cite{jd_paper}. 

\section{Quantized alternating current graphene strip \label{sec:periodic_ribbon}}

\begin{figure}
\includegraphics[width=\columnwidth]{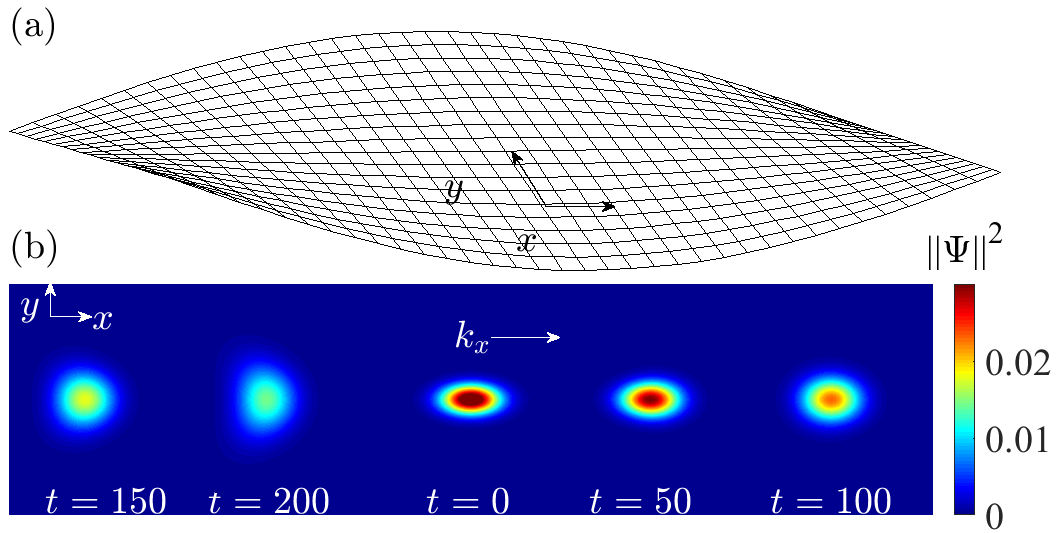}
\caption{\label{fig:periodic_schematic} \textbf{(a)} Real space geometry of the graphene strip, $x,y$ denote the coordinate directions. \textbf{(b)} Density plots of the wave-packet for different time-steps, the $k_x$ arrow denotes the propagation direction. The bulk of the wave-packet propagates forward and is spreading as expected. 
} 
\end{figure}

To investigate the potential of curvature on curved graphene sheets, we propose a periodic 
system with alternating current (AC) behaviour, which is quantized according to its shape.
The system geometry is initialized by the discrete mapping (or chart), 
  \begin{align}
    h^\alpha(x,y)	= & \begin{pmatrix}
           x \\
           y \\
           y \sin(\eta x/2)
         \end{pmatrix}
  \end{align}
with $x \in \{0, 2\pi \}$, $ y \in \{-L_y/2, L_y/2 \}$, $L_y$ being the domain size in the $y$ dimension, see Fig.~\ref{fig:periodic_schematic}. 
The boundaries are periodic along the $x$-direction and closed at $-L_y/2, L_y/2$.

The initial condition is given by a Gaussian wave-packet of the form:
  \begin{align}
    \Psi(\mathbf{r},\mathbf{k})= \frac{1}{\sqrt{2 \pi \mathcal{\s}^2}} & \begin{pmatrix}
           1 \\
           \lambda e^{i\theta}
         \end{pmatrix}
         e^{-\frac{|\mathbf{r}|^2}{4\mathcal{\s}^2}+i\mathbf{k}\cdot\mathbf{r}},
  \end{align}
where $\lambda=\pm 1$ is the band index, $\theta=\arctan(k_y/k_x)$, $\mathcal{\s}$ is a measure 
of the width, $\mathbf{r}=(x,y)$, $x,~y$ are the two space coordinates and $\mathbf{k}=(k_x,k_y)$, $k_x,~ k_y$ 
represent the $x$ and $y$ momenta, respectively. 

The initial values are taken as $k_x=1$, $k_y=0$ and $\lambda=1$. 
In the simulations, we consider a rectangular sheet with periodic boundary conditions 
on a grid of size $L_x \times L_y= 256 \times 128$ or $20nm \times 5nm$, while the external 
potential $A_a$ is set to zero. 
Therefore, the subsequent motion is purely curvature-driven.

The discretization of the real space shape of the graphene strip, is plotted in Fig.~\ref{fig:periodic_schematic}(a).
The norm of the wave-function, $\norm{\Psi}$, i.e. the probability density, $\rho$ is plotted 
in Fig.~\ref{fig:periodic_schematic}(b) for the initial and a few subsequent time-steps. 

As one can appreciate, the wave-packet spreads as expected, with no clear indication of motion along the $y$ direction.
  
\begin{figure}[!ht]
\includegraphics[width=\columnwidth, height=\columnwidth]{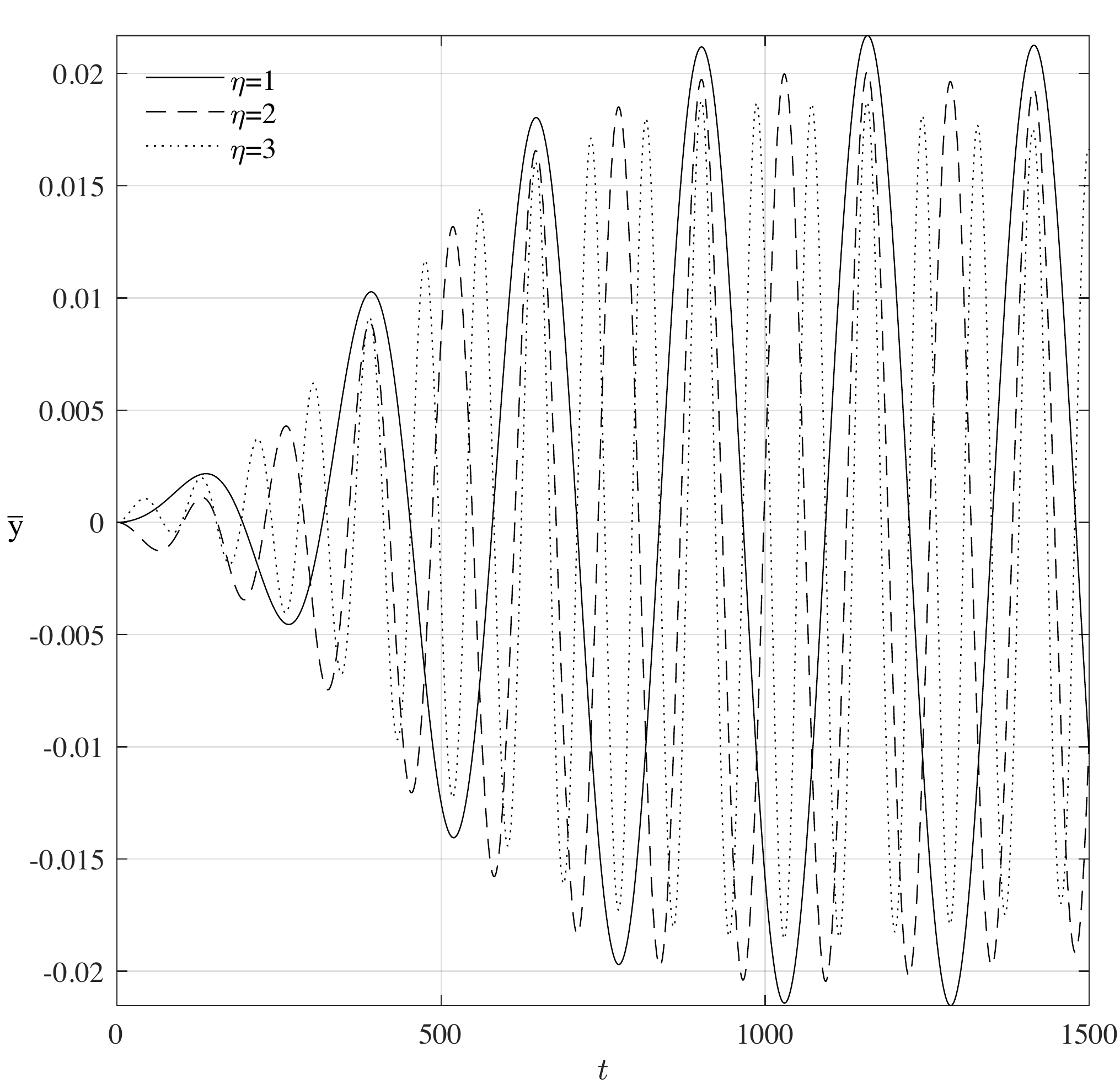}
\caption{\label{fig:periodic_main} Time evolution of relative position of the center of charge in the radial direction for geometries with $\eta=1,2,3$.}
\end{figure}

The position of the center of charge density along the $y$ direction: 
\begin{equation}
\bar{y}=\bigg( \int^{area} y (\rho(t) - \rho(0)) dA \bigg) /\int^{area} \rho(t) dA, 
\end{equation}
is plotted as a function of time in Fig.~\ref{fig:periodic_main}, where $dA=dxdy$.
 
A small but significant oscillation along the $y$ direction is observed.

These oscillations can be understood as the geometrical equivalent of the Bloch oscillations 
and they are a consequence of the sinusoidal, periodic domain, with the frequency 
quantized in units of the parameter $\eta$. 
For a slowly perturbed Hamiltonian and expanding around the wave-packet center
 $c$ (initialized to $(0,0)$ here), $H=H_c+\Delta H$, assuming 
a periodic system described by a Bloch wavefunction, the semi-classical equations of motion 
are given by \cite{berry_paper}:
\begin{align}
 \dot{\mathbf{r}}_c &=\frac{\del \e}{\del \mathbf{k}} - ( \Omega_{\mathbf{k}\mathbf{r}}\cdot \dot{\mathbf{r}}_c + \Omega_{\mathbf{k}\mathbf{k}}\cdot \dot{\mathbf{k}}_c)-\Omega_{\mathbf{k}t} \label{eq:semiclassical1} \\
  \dot{\mathbf{k}}_c &=\frac{\del \e}{\del \mathbf{r}} - ( \Omega_{\mathbf{r}\mathbf{k}}\cdot \dot{\mathbf{k}}_c + \Omega_{\mathbf{r}\mathbf{r}}\cdot \dot{\mathbf{r}}_c)-\Omega_{\mathbf{r}t}\label{eq:semiclassical2} 
\end{align}
where $\mathbf{r}_c, \mathbf{k}_c$ are the center of mass position and momentum of the wavepacket, $\mathbf{r}, \mathbf{k}$ are the position and momentum vectors, $t$ is time,
$\e$ is the band energy and $\Omega_{\mathbf{k}\mathbf{r}}=(\Omega_{\mathbf{k}\mathbf{r}})_{\a\b}= \del \mathbf{k}_\a \mathcal{A}_{\mathbf{r}_\b}- \del \mathbf{r}_\b \mathcal{A}_{\mathbf{k}_\a} $ is the Berry curvature and $\mathcal{A}_{\mathbf{r}}$ the Berry connection. 

As shown in the Appendix~\ref{app:berry_phase}, the Berry phase, and thus Berry curvature, can 
be related directly to the spin connection $\G_\mu$ through
$\mathcal{A}^i_{n}(\mathbf{R})=i\langle \Psi(\mathbf{R})| \del_\mathbf{R}|\Psi(\mathbf{R}) \rangle \implies \mathcal{A}_{R_i}=\tr \langle \Psi|\Gamma_i|\Psi \rangle$ for some parameter space $\mathbf{R}$ and eigen-function index $n$. 

The non zero terms of Eqs.~(\ref{eq:semiclassical1},\ref{eq:semiclassical2}) for the specific geometry   are $  \dot{\mathbf{r}}_c=\del \e / \del \mathbf{k}\approx v_f, ~ \dot{\mathbf{k}}_c=\Omega_{\mathbf{r}\mathbf{r}} \cdot \dot{\mathbf{r}}_c$, which imply:
\begin{equation}
    \frac{\del \mathbf{k}_\a}{\del \mathbf{r}_\beta}= (\Omega_{\mathbf{r}\mathbf{r}})_{\a\b}.
\end{equation}
For small-amplitude local wave-packets:
\begin{equation}
\delta \mathbf{k}_\a = \int (\Omega_{\mathbf{r}\mathbf{r}})_{\a\b} dr_\b=\int (\del \mathbf{r}_\a \mathcal{A}_{\mathbf{r}_\b}- \del \mathbf{r}_\b \mathcal{A}_{\mathbf{r}_\a} )d\mathbf{r}_\b
\end{equation}
and thus $\delta \mathbf{k}_y\propto sin(\eta x)$  and $\delta \mathbf{k}_x \propto cos(\eta x)$.

Therefore, the oscillations can be explained in terms of a real space Berry curvature, 
jointly with the classical geodesic equation on the 
corresponding manifold. 

The frequency of these Bloch-like oscillations is quantized according to $\eta$. Finally, some forward moving charge, even if driven, will experience an equivalent transverse oscillating motion, therefore the system might be implemented as a periodic, quantized oscillating current device.

\section{Conclusions and Outlook}

We have proposed the realization of a quantized alternating current on a curved graphene sheet. 
The oscillating current is numerically computed through the quantum lattice Boltzmann method 
in curved space and verified analytically via a set of semi-classical equations relating 
the Berry curvature to real space curvature. 
We interpret this result as a geometrical analogue of the Bloch oscillations, quantized according 
to the geometrical period $\eta$.

Building on these results, more complex and adjustable graphene devices can be envisaged 
in the context of curvature-based design. For example, the proposed quantized oscillating 
current on graphene, in conjunction with trapping quantum dots \cite{cdp}, could 
form the building  block for quantum information processing algorithms.

\begin{acknowledgments}
KF and HJH are grateful for the financial support Swiss National Science Foundation, under 
Grant No. 200021 165497 and to FUNCAP and CAPES for support.
SS acknowledges funding from the European Research Council under the European
Union Horizon 2020 Framework Programme (No. FP 2014-2020) ERC Grant Agreement No. 739964 (COPMAT).
\end{acknowledgments}

\bibliographystyle{ieeetr}
\bibliography{allcitations.bib}


\appendix
\section{Curved-space quantum lattice Boltzmann \label{sec:QLBM}}

The quantum lattice Bolzmann (QLB) method used for solving the Dirac equation as minimally coupled to curved space is an extension of the original method developed by Succi et al.  \cite{succi_qlbm}. The method exploits the conceptual similarities between the Dirac equation and the Boltzmann equation on the lattice.  
We present here the QLB method for a three-dimensional manifold, with straight forward usage to lower dimensional systems, \cite{Dellar_convqlbm,Dellar_isotropyqlbm,p_qlbmreview}. 

\subsection{The Dirac equation}

The classical Boltzmann equation for a particle density distribution function $f(x_a, v_a, t)$ is given by
\begin{equation}
\del_t f + v^i \del_{x^i}f=\mathcal{C}[f]-F^a\del_{v^a}f,
\end{equation}
 the left-hand side describes the advection of the distribution function, velocity $v^a$, whereas 
 the right-hand side describes the collisions between particles and the effect of external forces $F^a$. 
 Furthermore, the Dirac equation in curved space in Eq.~\ref{eq:Dirac} can be cast 
 into a kinetic theory form,
\begin{equation}
\label{eq:dirac_qlbm}
\del_t \Psi + \s^a\del_a \Psi = \mathcal{C} \Psi + \mathcal{F} \psi.
\end{equation}
Therefore, similarly to the Boltzmann equation,  the left hand side represents the 
'free streaming' step along matrix valued 'velocities' $\s^i$ while the right hand site 
contains a 'Collision' and a 'Forcing' term.

The collision term of Eq.~\ref{eq:dirac_qlbm} is represented by
\begin{equation}
\label{eq:collision}
\mathcal{C}=-(i m \g^0 + \s^a e_a^i\G_i), 
\end{equation}
where $m$ is the fermion mass.
The 'forcing term' is given by:
\begin{equation}
\label{eq:forcing_qlbm}
\mathcal{F}=-\s^a(e_a^i-\d_a^i) \del_i.
\end{equation}
where the symbols have their usual meaning. 
The partial derivative of the Dirac equation is distributed between the streaming part and the 
forcing term, resulting in a lattice-compatible classical streaming operator of 
the form $\del_t + v^a\del_a$, where $v^a \in \mathbb{Z}$. 
The forcing term is a consequence of the generalized Dirac matrices $\g^i=e_a^{~i}\g^a$ and captures the bulk of the curvature effects. The partial derivative in Eq.~\ref{eq:forcing_qlbm} is approximated by a local lattice finite difference scheme .    

\subsection{Diagonal streaming operator}

In order to obtain a diagonal streaming operator the complex $\s$-matrices have to be diagonalized first, which 
yields a diagonal velocity matrix with eigenvalues $v^a=\pm 1$. 
The diagonalization is achieved by suitable "rotation matrices":

\begin{align*}
	X_a^\dagger \,\s^a\, X_a  
	= \begin{pmatrix}
			 1 & 0 & 0 & 0 \\
			 0 & 1 & 0 & 0 \\
			 0 & 0 & -1 & 0 \\
			 0 & 0 & 0 & -1
		\end{pmatrix}
	= \g^0 \qquad \text{for } a=0,1,2,
\end{align*}
where the unitary transformation matrices $X_1, X_2, X_3$ are given by:

\begin{align*}
   \setlength{\arraycolsep}{2pt}
   \renewcommand{\arraystretch}{0.8}
	X_1 &= \T\frac{1}{\sqrt 2} 
		\begin{pmatrix}
			 1 & 0 & -1 & 0 \\
			 0 & 1 & 0 & -1 \\
			 0 & 1 & 0 & 1 \\
			 1 & 0 & 1 & 0
		\end{pmatrix},
	~
	X_2 = \T\frac{1}{\sqrt 2} 
		\begin{pmatrix}
			 0 & i & 0 & 1 \\
			 -i & 0 & i & 0 \\
			 -1 & 0 & -1 & 0 \\
			 0 & -1 & 0 & -i
		\end{pmatrix},
		\\
        X_3 &= \T\frac{1}{\sqrt 2} 
		\begin{pmatrix}
			 1 & 0 & 0 & -1 \\
			 0 & 1 & 1 & 0 \\
			 1 & 0 & 0 & -1 \\
			 0 & 1 & 1 & 0
		\end{pmatrix}.
\end{align*}

The streaming and collision operations are performed in successive steps using operator splitting, since 
the simultaneous diagonalization of the three $\s$ matrices is not possible:
 \begin{align*}
	\Psi(t+\frac{\dt}{D}) &=
	\exp\big(-\dt\s^1\del_1+\frac{\dt}{D}(\mathcal{C}+\mathcal{F})\big) \Psi(t), \\
	\Psi(t+\frac{2\dt}{D}) &=
	\exp\big(-\dt\s^2\del_2+\frac{\dt}{D}(\mathcal{C}+\mathcal{F})\big) \Psi(t+\frac{\dt}{D}), \\
		\Psi(t+\dt) &=
	\exp\big(-\dt\s^3\del_3+\frac{\dt}{D}(\mathcal{C}+\mathcal{F})\big) \Psi(t+\frac{2\dt}{D}),
\end{align*}
where $D=3$ denotes the spatial dimensions. 
Each streaming step can be diagonalized by left-multiplying with $X_a^\dagger$.
 \begin{align}
 \label{eq:diagonalization}
X_a^\dagger	\Psi(t+\frac{\dt}{D}) &=
	\exp\big(-\dt\s^a\del_a+\dt(\tilde{\mathcal{C}}_a+\tilde{\mathcal{F}}_a)\big) \tilde{\Psi}_a(t),
\end{align}
with the definitions:
 \begin{align*}
\Tilde{\Psi}_a \coloneqq X_a^\dagger \Psi, ~~ \tilde{\mathcal{F}}_a \coloneqq \frac{1}{2}X_a^\dagger \mathcal{F} X_a, ~~ \tilde{\mathcal{C}}_a \coloneqq \frac{1}{2} X_a^\dagger \mathcal{C} X_a,
\end{align*}
for $a=1,2,3$ (no Einstein summation is used here).
The exponential approximated as
 \begin{align*}
& \exp\big(-\dt\s^a\del_a+ \dt(\tilde{\mathcal{C}}+\tilde{\mathcal{F}})\big) \\
& \approx \big(\mathbb{1} -\dt\s^a \del_a+\dt(\tilde{\mathcal{C}_a}+\dt\tilde{\mathcal{F}}_a)+
\\
&+(\mathbb{1} - \frac{\dt}{2}\tilde{\mathcal{C}}_a)^{-1}(\mathbb{1} +\frac{\dt}{2}\tilde{\mathcal{C}}_a)\big)
\end{align*}
The expansion of the collision operator $e^{\dt\tilde{\mathcal{C}}_a}$ is unitary and thus conserves exactly the probability of the wavefunction. The streaming $e^{-\dt\g^0\del_a}$ and forcing $e^{\dt\tilde{\mathcal{F}}_a}$ operators are not expanded, as this is prohibited by the derivative. A simple $2^{nd}$-order expansion is performed, limiting the probability norm to $\Delta t^2$ accuracy. 
The operator splitting implies an error of order $\mathcal{O}(\dt^2)$, as 
$e^{\dt X}\cdot e^{\dt Y}=e^{\dt(X+Y)+1/2\dt^2[X,Y]}=e^{\dt(X+Y)}+\mathcal{O}(\dt^2)$.

The manifold is described by a chart $h$ defined in linear space, discretized on a regular rectangular lattice. 
The curved space quantum Lattice Boltzmann method evolves the four-spinor $\Psi = (\Psi^+, \Psi^-) = (\Psi_1^+,\Psi_2^-,\Psi_1^-,\Psi_2^+)$ from $t$ to $t+\delta t$. 
Once the operators are split, the following algorithm is performed in sequence for each 
lattice direction $n_a$, where $n_1=(1,0)$, $n_2=(0,1)$ and $a=1,2$. 

\begin{enumerate}
\item \textbf{Rotation:} The spinor is rotated by $X_a$ 
\begin{equation}
\tilde{\Psi}_a(x,t)=X^\dagger_a \Psi(x,t).
\end{equation}

\item \textbf{Collisions and curvature:} The collision and force operators are applied to the rotated spinor,
\begin{equation*}
\tilde{\Psi}_a^*(x,t)=\big( \Delta t \tilde{\mathcal{F}}_a + ( \mathbb{1} - \frac{\Delta t}{2} \tilde{\mathcal{C}}_a)^{-1}  ( \mathbb{1} + \frac{\Delta t}{2} \tilde{\mathcal{C}}_a)   \big) \tilde{\Psi}_a(x,t),
\end{equation*}
where $\tilde{\Psi}_a^*(x,t)$ denotes an auxiliary field,
\begin{equation}
\label{eq:collisiontilde}
\tilde{\mathcal{C}}_a=\frac{1}{2}X^\dagger_a \mathcal{C}X_a=-\frac{i}{D}m(X^\dagger_a \gamma^0 X_a)- \gamma^0e_a^i\Gamma_i,
\end{equation}

\begin{equation}
\label{eq:forcingtilde}
\tilde{\mathcal{F}}_a\tilde{\Psi}_a(x,t)=\big(e_a^i-\delta_a^i\big) \Big(\tilde{\Psi}_a(x \mp n_i \Delta t,t)-\tilde{\Psi}_a(x,t) \Big),
\end{equation}

where $n_i$ is the lattice direction and $\mathcal{C}$ is the collision term, Eq.~(\ref{eq:collision}).
The upper sign applies to the spin-up components $(\Psi_1^+,\Psi_2^+)$ and the lower sign to the spin-down 
ones $(\Psi_1^-,\Psi_2^-)$.
\item \textbf{Streaming:} The spinor components are streamed to the closest grid points along the lattice direction $\pm n_a$,
\begin{equation}
\tilde{\Psi}_a(x,t+\frac{\Delta t}{2})=\tilde{\Psi}_a^*(x \mp n_a\Delta t,t).
\end{equation}

\item \textbf{Inverse Rotation:} The spinor is rotated back via $X_a$,
\begin{equation}
\Psi_a(x,t+\frac{\Delta t}{2})=X_a \tilde{\Psi}_a(x,t+\frac{\Delta t}{2}).
\end{equation}

\item Repeat steps 2-4 for the next spatial direction
\end{enumerate}

 The external potentials $V(x)$, scalar, and $A(x)$, vector are added to the collision operator Eq.~\ref{eq:collisiontilde}, such that
\begin{equation}
\tilde{\mathcal{C}}_a=\frac{1}{2}X^\dagger_a \mathcal{C}X_a=-\frac{i}{D}(m-V)(X^\dagger_a \gamma^0 X_a)- \gamma^0e_a^i(\Gamma_i-iA_i).
\end{equation}
 
The simulation for strained graphene is carried out with modified Eqs.~(\ref{eq:collisiontilde},\ref{eq:forcingtilde}), according
to the following scheme:  
\begin{equation*}
\tilde{\mathcal{C}}_a \rightarrow \sqrt{g} \tilde{\mathcal{C}}_a, ~ e_a^i\rightarrow \sqrt{g}e_a^i.
\end{equation*}
The additional factor $\sqrt{g}$ originates from the volume element of the Hamiltonian Eq.~(\ref{eq:hamiltoniangraphene}).

\section{Berry phase relation to the spin connection \label{app:berry_phase}}


To solve the Dirac equation, minimally coupled to curvature, Eq.~\ref{eq:Dirac}, with $A_i=\mathbf{0}$ and assuming that the wave-packet has a negligible spread, $\delta \mathbf{r} \rightarrow 0$, the connection component 
of the covariant derivative can be absorbed into the wavefunction,  so that
\begin{equation}
\label{eq:wavefunction}
    \Psi \rightarrow \Psi \exp\bigg(i \int_{\mathbf{r}_c}^{\mathbf{r}_c+\delta \mathbf{r}} \Gamma_i d\mathbf{r} \bigg) 
\end{equation}
where $\mathbf{r}_c$ is the center of mass position and $\Gamma_i$ is the spin-connection matrix. 
For a Gaussian wavepacket with spread $\sigma$ and momentum $\mathbf{k}$, the 
wavefunction Eq.~ (\ref{eq:wavefunction}) takes the following form
\begin{equation}
     \Psi(\mathbf{r},\mathbf{k})   = \frac{1}{\sqrt{2 \pi \mathcal{\s}^2}} 
     \begin{pmatrix}
           1 \\
           0 \\
           0 \\
           -1
         \end{pmatrix}
     e^{i \int \Gamma_i d\mathbf{r}} e^{-\frac{|\mathbf{r}|^2}{4\mathcal{\s}^2}+i\mathbf{k}\cdot\mathbf{r}}.
\end{equation}
 This wave-function minimally couples the standard Dirac equation to curved space through the spin connection.
Defining the Berry connection as 
\begin{equation}
\label{eq:berry_conection}
    \mathcal{A}^i_n(\mathbf{R})=i\langle \Psi(\mathbf{R})| \del_\mathbf{R} | \Psi(\mathbf{R}) \rangle
\end{equation}
for some parameter space $\mathbf{R}$ and eigen-function $n$. 
The Berry phase can be calculated from the complete loop integral of the connection, acoording to:
\begin{equation}
    \gamma=\oint_0^{2 \pi} \mathcal{A}(\mathbf{R})g^{i j}_{\mathbf{R}}d\mathbf{R}.
\end{equation}
In a similar manner to the treatment of the Aharonov-Bohm effect from Berry \cite{berry_paper}, we define 
the slow and fast coordinates $R$ and $r$ respectively, such that $\Psi(R,r) \rightarrow \Psi(r-R)$. 
The wave-function takes then the  form
\begin{equation}
     \Psi_r(R-r)   = \frac{1}{\sqrt{2 \pi \mathcal{\s}^2}} 
     \begin{pmatrix}
           1 \\
           0 \\
           0 \\
           -1
         \end{pmatrix}
     e^{i \int \Gamma_r d r} e^{-\frac{|R-r|^2}{4\mathcal{\s}^2}+ik(R-r)}.
\end{equation}
From Eq.~\ref{eq:berry_conection} the explicit form of the wavefunction implies that $\mathcal{A}^i=\tr \Gamma_i$. 
The implication of these result is that the Berry connection and curvatures can be directly related to the 
real space affine connection and Ricci curvature tensor under suitable conditions.   

As a consequence, the phase change of a wavepacket moving around a closed loop, can be calculated 
from the Berry phase. Integrating naively around a closed loop
\begin{equation}
    \gamma=\oint_0^{2 \pi} \tr \langle \Psi_r| \del_r  \Psi_r \rangle g^{1 1} d\mathbf{r},
\end{equation}
where $\tr$ denotes the trace of the resulting matrix and takes into account 
the spinorial character of the Dirac wavefunction.
\end{document}